\documentstyle[11pt]{article} \baselineskip=18pt
\newcommand{\slp}{\raise.15ex\hbox{$/$}\kern-.57em\hbox{$\partial$}}
\newcommand{\sla}{\raise.15ex\hbox{$/$}\kern-.57em\hbox{$a$}}
\newcommand{\slA}{\raise.15ex\hbox{$/$}\kern-.57em\hbox{$A$}}
\newcommand{\slB}{\raise.15ex\hbox{$/$}\kern-.57em\hbox{$B$}}
\newcommand{\slb}{\raise.15ex\hbox{$/$}\kern-.57em\hbox{$b$}}
\newcommand{\slW}{\raise.15ex\hbox{$/$}\kern-.57em\hbox{$W$}}
\newcommand{\be}{\begin{equation}} \newcommand{\ee}{\end{equation}}
\newcommand{\bear}{\begin{eqnarray}} \newcommand{\ear}{\end{eqnarray}}

 \newcommand{\Ha}{\cal H} 

\begin{document}

\title{BRST Cohomology and Hilbert Spaces of Non-Abelian Models in the
Decoupled Path Integral Formulation} \author{ \sc{K.D. Rothe$^a$,F.G.
Scholtz$^b$ and A.N. Theron$^b$ }\\ \sc{$^a$  \small{\it Institut f\"ur
Theoretische Physik, Universit\"at Heidelberg,}}\\ \sc{\small{\it
Philosophenweg 16, 69120 Heidelberg, Germany}}\\ \sc{$^b$ \small{\it Institute
of Theoretical Physics, University of Stellenbosch,}}\\ \sc{\small{\it
Stellenbosch 7600, South Africa}}}

\date{17 September 1996}

\maketitle

\begin{abstract}
The existence of several nilpotent Noether charges in the decoupled
formulation of two-dimensional gauge theories does not imply that all
of these are required to annihilate the physical states.  We elucidate
this matter in the context of simple quantum mechanical and field
theoretical models, where the structure of the Hilbert space is known.
We provide a systematic procedure for deciding which of the BRST
conditions is to be imposed on the physical states in order to ensure
the equivalence of the decoupled formulation to the original,
coupled one.
\end{abstract}

\newpage

\section{Introduction}

Bosonization techniques have proven useful in solving two--dimensional quantum
field theories.  In particular the Schwinger and Thirring models have been
solved in this way \cite{2}.  In the path integral framework the solubility
of the
Schwinger model manifests itself in the factorization of the partition function
in terms of a free massive positive metric field, a free negative metric zero
mass field and free massless fermions \cite{2,3}.  It has long been
realized that this
factorization can be understood as a chiral change of variables in the path
integral \cite{4}.  In this decoupled formulation the physical Hilbert
space of gauge
invariant observables of the original model is recovered by implementing the
BRST conditions associated with the usual gauge fixing procedure and the
chiral change of variables \cite{8}.  These conditions are identical to
those originally obtained by Lowenstein and Swieca \cite{5} on the operator
level, stating that the physical Hilbert space be annihilated by the sum of the
currents of the negative metric fields and the free fermions.

Similar bosonization techniques have been applied to Quantum Chromodynamics in
1+1--dimensions (QCD$_2$).   A review can be found in \cite{2,6}.
 Analogous to the case of the
Schwinger model,
it has recently been shown that for a suitable choice of integration variables
the partition function of QCD$_2$ factorizes in terms of free massless
fermions, ghosts, negative level Wess Zumino Witten fields and fields
describing massive degrees of freedom \cite{7}.  In the case of one flavor
the sector
corresponding to the massless fields was found \cite{8} to be equivalent to
that of a
conformally invariant, topological $G/G$ coset model \cite{9} with $G$ the
relevant gauge group.  As a result of the
gauge fixing and the decoupling procedure
there were shown to exist several nilpotent charges \cite{13}
associated with BRST-like symmetries.  These charges were found to be second
class.  This raises the question as to whether all or just some of these
charges are required to annihilate the physical states.

 Assuming the ground state(s)
of $QCD_2$ to be given by the state(s) of the conformally invariant sector,
 (as is the case in the Schwinger model), the solution of the
corresponding cohomology problem led to the conclusion \cite{8} that the ground
state of
$QCD_2$ with gauge group $SU(2)$ and one flavor is 2 times twofold degenerate,
and
corresponding to the primaries of the $(U(1) \times SU(2)_1)/SU(2)_1$ coset
describing
the conformal sector (with $U(1)$ playing a spectator role). Since there
are however also
BRST constraints linking the coset-sector to the sector of massive
excitations, the above
hypothesis is not necessarily realized.

  In ref. \cite{10}
the idea of smooth bosonization was introduced whereby
two dimensional path integral bosonization is formulated in terms of a
  gauge fixing procedure. To accomplish this, a ``bosonization''
gauge symmetry with an associated ``bosonization''
BRST symmetry were introduced.
It was argued in ref. \cite{11} that the most natural choice of gauge to
recover the canonical bosonization dictionary is to ``gauge fix'' the fermions
in a $U(N)/U(N)$ coset model.  In fact, non-abelian smooth bosonization has
only been achieved  by this choice of ``gauge''  \cite{11,12}.  The main result
of this approach to bosonization \cite{11} is that the free fermion partition
function factorizes into the partition function of a $U(N)/U(N)$ coset model
and a WZW model. An interpretation of this result along the lines mentioned
above for the case of $QCD_2$ would lead one to conclude that the spectrum of
free $U(2)$ fermions is two-fold degenerate.  This conclusion is, however,
wrong since the ``bosonization'' BRST links the $U(N)/U(N)$ coset model to the
"matter" sector described by the WZW model \cite{11} (see section 3).
Therefore the BRST constraints play an essential role in identifying the
correct spectrum.

The above examples illustrate that extreme care has to be taken with regard
to the
implementation of the BRST
symmetries when identifying the physical states.   In particular the
identification of  the BRST symmetries
on the decoupled level  can be misleading as not all the charges
 associated with these symmetries are generally
required to annihilate the physical states in order to ensure
 equivalence with the original, coupled
formulation.   In section 2 we illustrate this point  using a simple
quantum mechanical model for which
the Hilbert space is known and the BRST symmetries and associated cohomology
 problem are very
transparent and easy to solve.  These considerations generalize to the field
theoretic case.  As examples
we discuss the non-abelian bosonization of free fermions  in section 3 and some
 aspects of QCD$_2$, as
considered in ref. \cite{13}, in section 4.

It is of course well known that the BRST symmetries on the decoupled level
 originate from the changes
of variables made to achieve the decoupling, as has been discussed in the
literature before (see e.g. ref.
\cite{14,24,25}).  The aspect we want to emphasize here is the role these
symmetries,
and the associated
cohomolgy,
play in constructing the physical subspace on the decoupled level, i.e., a
subspace
isomorphic to the
Hilbert space of the original coupled formulation.   In particular we want
to stress
that the mere existence
of  a nilpotent symmetry does not imply that the associated charge must
annihilate the
 physical  states.   We
are therefore seeking a procedure to decide on the latter issue.  We
formulate such
a procedure in section 2
and apply it in sections 3 and 4.  It amounts to implementing the changes
of variables
 by inserting an
appropriate identity in terms of auxiliary fields in the functional
integral.  BRST
transformation rules for
these auxiliary fields are then introduced such that the identity amounts
to the
addition of a $Q$-exact
form
to the action.  The dynamics is therefore unaltered on the subspace of states
annihilated by the
corresponding BRST charge.  We must therefore require that this charge
annihilate
the physical states.
The auxiliary fields are then systematically integrated out and the BRST
transformation rules "followed"
through the use of equations of motion.   This simple procedure allows for the
identification of the BRST
charges on the decoupled level which are required to annihilate the physical
subspace isomorphic to the
Hilbert space of the coupled formulation.

The paper is  organized as follows: In section 2 we illustrate the points
raised above in terms of a
simple quantum mechanical model.  Using this model as example we also
formulate a
 general procedure
to identify the BRST charges  required to annihilate the physical states
such that
 equivalence with the
coupled formulation is ensured on the physical subspace.  In sections 3 and 4
 we apply this procedure to the non-abelian bosonization of free Dirac fermions
in 1+1 dimensions and to QCD$_2$, respectively.

\section{Quantum mechanics}

In this section we discuss a simple quantum mechanical model to illustrate
the  points raised in the introduction.  Since
the emphasis is on the structures of the Hibert spaces on the coupled
and decoupled levels, and the role which the
BRST symmetries and  associated cohomology play in this  respect,
 we introduce the
 model  in second quantized form
and only then set up a path integral formulation  using
coherent states. The model is then bosonized (decoupled)
and the BRST symmetries and cohomologies are
 discussed.

Consider the model with Hamiltonian

$$\hat{H} = 2g \, \hat{J} \cdot \hat{J} \eqno{(2.1)}$$
where $\hat{J}^a$ are the generators of a $SU(2)$ algebra.  We realize this
algebra on Fermion Fock space in the following way \cite{15}:

$$\begin{array}{rcl} \hat{J}^+ &=& \displaystyle \sum_{m>0} \, a^\dagger_m \,
a^\dagger_{-m} \, , \\[7mm] \hat{J}^- &=& \displaystyle \sum_{m>0} \, a_{-m} \,
a_m \, , \\[7mm] \hat{J}^0 &=& \frac{1}{2} \displaystyle \sum_{m>0} \,
(a^\dagger_m \, a_m - a_{-m} \, a^\dagger_{-m}) \, . \end{array}
\eqno{(2.2)}$$ Here $a^\dagger_m \, (a_m)$, $|m| = 1, 2, \dots N_f$ are
fermion creation (annihilation) operators.  The operators (2.2) provide a
reducible representation of the usual commutation relations of angular momentum
and the representations carried by Fermion Fock space are well known \cite{16}.
The index $|m|$ plays the role of flavor with $N_f$ the number of flavors.
The Hamiltonian (2.1) thus describes a $SU(2)$ invariant theory with $N_f$
flavors, as will become clear in the Lagrange formulation discussed below.

The spectrum of the Hamiltonian $\hat{H}$ is completely known,
the eigenvalues of $\hat{H}$ being given by

$$E_j = 2 \, g \, j \, (j + 1) \eqno{(2.3)}$$
where each eigenvalue is $g_j \, (2j + 1)$-fold degenerate, with $g_j$ the
number
of times the corresponding irreducible representation occurs. In particular,
for $N_f=1$ the spectrum of $\hat{H}$ consists of a doublet, as well as two
singlets describing a two-fold degenerate state of energy $E=0$. For
positive $g$
this corresponds to a two-fold degenerate ground state.

We express the  vacuum--to--vacuum amplitude associated with $\hat{H}$
as a functional integral over Grassmann
variables.  For this purpose introduce the fermionic coherent state \cite{17}

$$| \, \chi > = \exp \, \left[- \frac{1}{2} \, \sum_{m>0} \, (\chi^\dagger_m \,
\chi_m + \chi^\dagger_{-m} \, \chi_{- m} + \chi_m \, a^\dagger_m + \chi_{-m} \,
a_{-m})\right] | \, 0 > \eqno{(2.4)}$$
where $\chi^\dagger_m$, $\chi_m$ are complex valued Grassmann variables and

$$a_m \, | \, 0 > = a_{-m}^\dagger \, | \, 0 > \, = 0 \, , \quad \forall m > 0
\, . \eqno{(2.5)}$$

We obtain the path integral representation of the vacuum--to--vacuum transition
amplitude by following the usual procedure \cite{18} and using the completeness
relation for the coherent states \cite{17}.  We find

$$Z = \int [d \eta^\dagger] \, [d \eta] \, e^{i \int \, dt \, L_F} \,
\eqno{(2.6{\rm a})}$$
where $L_F$ is the "fermionic" Lagrangian

$$L_F = \eta_f^\dagger \, (i \, \partial_t + m) \, \eta_f - g \, {\rm tr} \,
j^2 \eqno{(2.6{\rm b})}$$
and a summation over the flavor index $f$  $(f = 1, 2, \dots N_f)$ is
implied. The
mass $m = - 3g$ arises from normal ordering with respect to the Fock vacuum
$| \, 0
>$ defined in (2.5).  Furthermore, $\eta_f$ denotes the two-component spinor

$$
\eta_f = \left(\begin{array}{c}
\chi_f\\
\chi_{-f}
\end{array}\right) \, , \eqno{(2.7{\rm a})}$$
\\[7mm]
and $j, j^a_f$ are the "currents"
$$\begin{array}{rcl}j &=& \displaystyle \sum_f \, j_f, \,
 j_f \; = \; j^a_f \, t^a \, , \\[7mm]
j^a_f &=& \eta^\dagger_f \, t^a \, \eta_f \end{array} \eqno{(2.7{\rm
b})}$$
where the $SU(2)$ generators are normalized as
${\rm tr} \, (t^a \, t^b) = \delta^{ab}$.

Introducing the field

$$B = B^a \, t^a \, , \eqno{(2.8)}$$
we can write the partition function as

$$
Z = \displaystyle \int \, [d \eta^\dagger] \, [d \eta] \, [dB]
 \, e^{i \int \, dt \, L} \eqno{ (2.9{\rm a})}
$$
$$
 L = \eta^\dagger_f \, (i \, \partial_t + B + m)
\, \eta_f + \frac{1}{2g} \, {\rm tr} \, B^2 \, \eqno{(2.9{\rm b})}
$$

The bosons and fermions can be decoupled by making the change of variables

$$B = V \, i \, \partial_t \, V^{-1} \eqno{(2.10{\rm a})}$$
where $V$ are group valued fields in the fundamental representation of $SU(2)$.
Simultaneously we make the change of variables

$$\psi_f = V^{-1} \, \eta_f \, . \eqno{(2.10{\rm b})}$$
The Jacobian associated with the transformation is (there are no anomalous
contributions in $0 + 1$ dimensions):

$$J = \det \, (D^{adj}_t \, (V)) = \det \, (\partial_t) \eqno{(2.11)}$$
where $D_t^{adj} \, (V)$ is the covariant derivative in the adjoint
representation

$$D^{adj}_t \, (V) = \partial_t + [V \, i \, \partial_t \, V^{-1} \, , ~] \, .
\eqno{(2.12)}$$
Representing the determinant in terms of ghosts we obtain for the partition
function the factorized form

$$Z = Z^{(0)}_F \, Z^{(0)}_{\rm gh} \, Z_V = \int \, [d \psi^\dagger] \, [d
\psi]
\, [d V] \, [db] \, [dc] \, e^{i \int \, dt \, L^{(0)}} \eqno{(2.13{\rm a})}$$
with

$$L^{(0)} = L^{(0)}_F + L^{(0)}_{\rm gh} + L_V \eqno{(2.13{\rm b})}$$
where

$$\begin{array}{rcl}
L^{(0)}_F &=& \psi^\dagger_f \, (i \, \partial_t + m) \,
\psi_f \, , \\[3mm] L^{(0)}_{\rm gh} &=& {\rm tr} \, b \, i \, \partial_t \, c
\, ,
\\[5mm] L_V &=& \frac{1}{2g} \, {\rm tr} \, (V \, i \, \partial_t \, V^{-1})^2
\, . \end{array}
\eqno{(2.13{\rm c})}$$
Here $b$ and $c$ are Lie algebra valued ghost fields $b = b^a \, t^a$ and
$c = c^a
\, t^a$.

In arriving at the above decoupled form of the partition function, we have not
mentioned the effect of the change of variables on the boundary of
the path integral.  It is important to realize that the decoupling
 of the partition function is not
affected by the implied change of the boundary condition since the
transformation (2.10) is local.

The Hilbert space associated with the factorized partition function is the
direct product of fermion, boson and ghost Fock spaces and is clearly much
larger than that of the original interacting model.  It is therefore natural to
ask what conditions should be imposed on this direct product space to recover
a subspace isomorphic to the original Hilbert space.  As these conditions are
of a group theoretical nature, it is necessary to first clarify the precise
content of these Hilbert spaces from a representation theory point of view
before the above mentioned isomorphism can be established.

As we have seen the states in the Hilbert space (Fermion Fock space) of the
original interacting model can be labeled by $| \, \alpha \, j \, m >$ where
$j$, $m$ labels the $SU(2)$-flavor representations and weights,
 respectively, and $\alpha$ is a multiplicity index.

On the decoupled level the Hilbert space is the direct product of fermion,
boson and ghost Fock spaces.  Upon canonical quantization it is again clear
that states in the (free) fermionic sector can be labeled by $| \, \alpha
\, j \, m
>_{\rm F}$ where the allowed values of $\alpha$ and $j$ coincide with those of
the interacting model.  The ghosts $c$ and $b$ are canonically conjugate
fields.  We take $b$ as the annihilation and $c$ as the creation operator.  The
ghost vacuum is then defined by $b^a \, | \, 0 \, > \, = 0$.  Defining the
ghost number operator by $N_{\rm gh} = c^a \, b^a \, , \,$
we see that $ b^a$ carries ghost
number $-1$ and $c^a$ ghost number 1.  Since we are interested in the physical
sector which is built on the ghost vacuum (ghost number zero), we do not need
to analyze the representation theory content of the ghost sector in more
detail.  It is therefore only the bosonic sector that requires a detailed
analysis.

Turning to the bosonic Lagrangian of (2.13c) we note the presence of a left
and right global symmetry

$$\begin{array}{rcl}
V &\longrightarrow& LV \, ,\\
V &\longrightarrow& VR
\end{array} \eqno{(2.14)}$$
where $L$ and $R$ are $SU(2)$ matrices in the fundamental representation
corresponding to
left- and right-transformations, respectively.
Following the Noether construction the conserved currents generating
these symmetries are identified as \cite{19}

$$\begin{array}{rcl} L^a &=& \frac{1}{g} \, {\rm tr} \, V \, i \, \partial_t
\, V^{-1} \, t^a \, ,\\[5mm] R^a &=& \frac{1}{g} \, {\rm tr} \, (i \,
\partial_t \, V^{-1}) \, V \, t^a \, , \end{array} \eqno{(2.15)}$$
respectively. In phase space this reads

$$\begin{array}{rcl}
L^a &=& {\rm tr} \, i \, V \tilde\pi_V \, t^a \, , \\[2mm]
R^a &=& {\rm tr} \, i \, \tilde\pi_V \, V \, t^a \, ,
\end{array} \eqno{(2.16)}$$
with $\pi_V$ the momentum canonically conjugate to $V$, and "tilde" denoting
"transpose".  The following Poisson brackets are easily verified

$$\begin{array}{rcl} \{L^a \, , \, L^b \}_{\rm P} &=& - f^{abc} \, L^c \,
,\\[2mm] \{R^a \, , \, R^b \}_{\rm P} &=& f^{abc} \, R^c \, ,\\[2mm] \{L^a \, ,
\, R^b \}_{\rm P} &=& 0 \, ,\\[2mm] \{L^a \, , \, V \}_{\rm P} &=& - i \, t^a
\, V \, ,\\[2mm] \{R^a \, , \, V \}_{\rm P} &=& - i \, V \, t^a \, ,
\end{array} \eqno{(2.17)}$$ where $f^{abc}$ are the $SU(2)$ structure
constants.

Canonical quantization proceeds as usual.  The Hilbert space of
this system is well known, and corresponds to that of the rigid rotator
\cite{19}.
Hence the Wigner $D$--functions $D^I_{MK}$ provide a realization in terms of
square integrable functions on the group manifold. It is important to note
that the Casimirs of the left and right symmetries both equal $I \, (I + 1)$.
Furthermore $M$ and $K$ label the weights of the left and right symmetries,
respectively.  To determine the allowed values of $I$ one notes from (2.17)
that $V$ transforms as the $j = \frac{1}{2}$ representation under left and
right transformations.  Thus $V$ acts as a tensor operator connecting
integer and
half--integer spins. It follows that $I$  can take the values
$I = 0, \frac{1}{2}, 1, \frac{3}{2} \dots$~.  We therefore conclude
that on the decoupled
level the states have the structure $| \, \alpha \, j \, m >_{\rm F} \, |
\, I \, M
\, K >_{\rm B} \, | \, gh >$ where the subscripts F, B refer
to the fermionic and bosonic sectors, respectively.  The allowed values
of the quantum numbers in these sectors are as discussed above.

Returning to the question as to which conditions are to be imposed on the
direct product space of the decoupled formalism in order
to recover the Hilbert space of
the original model, we note by inspection of (2.13) the existence of three
BRST symmetries (of which only two are independent).  One of them acts in all
three sectors and is given by

$$\begin{array}{rcl}
\delta_1 \, \psi &=& c \, \psi \, , \\[2mm]
\delta_1 \, \psi^\dagger &=& \psi^\dagger \, c \, , \\[2mm]
\delta_1 \, V &=& - V \, c \, , \\[2mm]
\delta_1 \, V^{-1} &=& c \, V^{-1} \, , \\[2mm]
\delta_1 \, b &=& - j^{(0)} - R + \{b, \, c\} \, \, , \\[2mm]
\delta_1 \, c &=& \frac{1}{2} \, \{c, \, c\} \, .
\end{array} \eqno{(2.18)}$$
Here $\delta_1$ is a variational derivative graded with respect to Grassmann
number, \{~\} denotes a matrix anti--commutator and

$$\begin{array}{rcl} j^{(0)} &=& \displaystyle \sum_f \, (\psi^\dagger_f \, t^a
\, \psi_f) \, t^a \, ,\\[5mm] R &=& R^a \, t^a \; = \; \frac{1}{g} \, (i \,
\partial_t \, V^{-1}) \, V \, . \end{array} \eqno{(2.19)}$$
The other BRST symmetries act in the fermion--ghost and boson--ghost sectors,
respectively, and are given by

$$\begin{array}{rcl}
\delta_2 \, \psi &=& c \, \psi \, ,\\[2mm]
\delta_2 \, \psi^\dagger &=& \psi^\dagger \, c \, ,\\[2mm]
\delta_2 \, b &=& - j^{(0)} + \{b \, , c\} \, ,\\[2mm]
\delta_2 \, c &=& \frac{1}{2} \, \{c \, , \, c\} \, .
\end{array} \eqno{(2.20)}$$
as well as

$$\begin{array}{rcl}
\delta_3 \, V &=& - V \, c \, ,\\[2mm]
\delta_3 \, V^{-1} &=& c \, V^{-1} \, ,\\[2mm]
\delta_3 \, b &=& - R + \{b \, , \, c\} \, ,\\[2mm]
\delta_3 \, c &=& \frac{1}{2} \, \{c \, , \, c\} \, .
\end{array} \eqno{(2.21)}$$
The above transformations are nilpotent.  Note, however, that they do
not commute.

Performing a canonical quantization, we define the ghost current

$$J_{\rm gh} = - : \, \{b \, , c\} \, : \eqno{(2.22)}$$
where : ~: denotes normal ordering with respect to the ghost vacuum.  The
nilpotent charges $Q_i$ generating the transformations (2.18) -- (2.21),
i.e., $\delta_i \, \phi = [Q_i \, , \phi]$, with $\phi$ a generic field and
[~,~] a graded commutator, have the general form:

$$\begin{array}{rcl}
Q_1 &=& - {\rm tr} \, [c \, (j^{(0)} + R + \frac{1}{2} \, J_{\rm gh})] \,
,\\[3mm]
Q_2 &=& - {\rm tr} \, [c \, (j^{(0)} + \frac{1}{2} \, J_{\rm gh})] \, ,\\[3mm]
Q_3 &=& - {\rm tr} \, [c \, (R + \frac{1}{2} \, J_{\rm gh})] \, .
\end{array} \eqno{(2.23)}$$

We remarked above that the direct product Hilbert space associated with the
facto\-rized form of the partition function is much larger than that of the
original interacting model.  We now inquire as to which of the above BRST
charges are required to vanish on the physical subspace (${\cal H}_{\rm ph}$)
in order to establish the isomorphism to the Hilbert space of the original
model.

We begin by showing that $Q_1$ is required to vanish on ${\cal H}_{\rm ph}$.
In order to illustrate the method, which will be used repeatedly, we briefly
sketch the main steps for the case at hand.  To implement the change of
variables (2.10a) we make use of the identity

$$1 = \int \, [dV] \, \delta \, (B - V \, i \, \partial_t \, V^{-1}) \, \det
\, i \, D^{adj}_t \, (V) \eqno{(2.24)}$$
where the covariant derivative was defined in (2.12).  Inserting this
identity into (2.9a), using the Fourier representation of the Dirac delta
and lifting the determinant by introducing Lie--algebra valued ghosts
$\tilde{b}$ and $\tilde{c}$ we have

$$Z = \displaystyle \int \, [d \, \eta^\dagger] \, [d \,
\eta] \, [dB] \, [d \, \lambda] \, [dV] \, [d \, \tilde{b}] \, [d \, \tilde{c}]
\, e^{i \int \, dt \, L'}\eqno{(2.25{\rm a})}$$
with
$$L' = L + \Delta L\eqno{(2.25{\rm b})}$$
where

$$\Delta L = {\rm tr} \, (\lambda \, (B - V \, i \, \partial_t \, V^{-1}) -
{\rm tr} \, (\tilde{b} \, i \, D^{adj}_t \, (V) \, \tilde{c}) \, .
\eqno{(2.25{\rm c})}$$
This Lagrangian is invariant under the BRST transformation

$$\begin{array}{rcl} \delta \, \eta &=& \delta \, \eta^\dagger \; = \; \delta
\, B \; = \; \delta \, \lambda \; = \; 0 \, ,\\[2mm] \delta \, \tilde{b} &=&
\lambda \, ,\\[2mm] \delta V &=& \tilde{c} \, V \, , \, \delta \, V^{-1} \; =
\; - V^{-1} \, \tilde{c} \, ,\\[2mm] \delta \, \tilde{c} &=& \frac{1}{2} \,
\{\tilde{c} \, , \, \tilde{c}\} \, . \end{array} \eqno{(2.26)}$$
One readily checks that this symmetry is nilpotent off--shell.

Noting that $\Delta L$ can be expressed as a BRST exact form

$$\Delta L = \delta \, (\tilde{b} \, (B - V \, i \, \partial_t \, V^{-1}))\,,
\eqno{(2.27)}$$
we conclude that equivalence with the original model is ensured on the subspace
of states annihilated by the corresponding BRST charge.

Next we show that the transformation (2.26) is in fact equivalent to the
transformation (2.18).  Using the equation of motion for $\lambda$ and $B$ we
obtain the BRST transformation rules:

$$\begin{array}{rcl}
\delta \psi &=& \delta \, \psi^\dagger \; = \; 0 \, ,\\[2mm]
\delta \, \tilde{b} &=& - \frac{1}{g} \, V \, i \, \partial_t \, V^{-1} - j
\, ,\\[5mm]
\delta \, \tilde{c} &=& \frac{1}{2} \, \{\tilde{c} \, , \, \tilde{c}\} \,
,\\[4mm]
\delta \, V &=& \tilde{c} \, V \, ,\\[2mm]
\delta \, V^{-1} &=& - V^{-1} \, \tilde{c} \, .
\end{array} \eqno{(2.28)}$$
One readily checks that this is a symmetry of the action with Lagrangian

$$L = \eta^\dagger_f \, (i \, \partial_t + V \, i \, \partial_t \, V^{-1} + m)
\, \eta_f + \frac{1}{2g} \, {\rm tr} \, (V \, i \, \partial_t \, V^{-1})^2 -
{\rm tr} \, (\tilde{b} \, i \, D^{adj}_t \, (V) \, \tilde{c}) \, .
\eqno{(2.29)}$$
obtained after integrating out $\lambda$ and $B$.
Finally we return to the decoupled partition function (2.13) by transforming
to the free fermions and ghosts

$$\begin{array}{rcl}
\psi_f &=& V^{-1} \, \eta_f \, ,\\[2mm]
c &=& - V^{-1} \, \tilde{c} \, V \, ,\\[2mm]
b &=& V^{-1} \, \tilde{b} \, V \, .
\end{array} \eqno{(2.30)}$$
In terms of these variables the BRST transformations (2.28) become those of
(2.18).  This demonstrates our above claim that the BRST charge generating
the transformation (2.18) has to vanish on ${\cal H}_{\rm ph}$ to ensure
equivalence with the original model.

An alternative way of proving the above statement is to note that the decoupled
Lagrangian $L^{(0)}$ of (2.13) can be expressed in terms of the original
fermionic Lagrangian $L_F$ of (2.6b) plus a $\delta_1$ exact part.  In terms
of the free fermions $ \psi_f$ and interacting fermions $\eta_f = V^{-1} \,
\psi_f$, we may rewrite $L^{(0)}$ as:

$$L^{(0)} = \eta^\dagger_f \, (i \, \partial_t + m) \, \eta_f + g \, {\rm tr}
\, (R j^{(0)}) + \frac{g}{2} \, {\rm tr} \, R^2 + {\rm tr} \, (b \, i \,
\partial_t \, c) \eqno{(2.31)}$$
which may be put in the form

$$L^{(0)} = \eta^\dagger_f \, (i \, \partial_t + m) \, \eta_f - \frac{g}{2} \,
{\rm tr} \, j^2 + \frac{1}{2} \, {\rm tr} \, (b \, i \, \partial_t \, c) -
\frac{g}{2} \, \delta_1 \, [{\rm tr} \, b \, (R + j^{(0)})] \, .
\eqno{(2.32)}$$
Comparing with (2.6) we see that $L^{(0)}$ and $L_F$ just differ by a BRST
exact term (up to a decoupled free ghost term).  Hence we recover the original
fermion dynamics on the sector which is annihilated by the BRST charge $Q_1$.
Therefore only the first of the three BRST symmetries (2.18) -- (2.21)
has to be imposed on the states.  To see what this implies, we now solve the
cohomology problem associated with $Q_1$.

As usual \cite{20} we solve the cohomology problem in the zero ghost number
sector
$b^ a \, | \, \Psi >_{\rm ph} \, = 0$.  The condition $Q_1 \, | \, \Psi >_{\rm
ph} \, = 0$ is then equivalent to (see (2.22) and (2.23))

$$(j^{(0)} + R) \, | \, \Psi >_{\rm ph} = 0 \, . \eqno{(2.33)}$$
The physical states $| \, \Psi >_{\rm ph}$ are thus singlets under the total
current $J = j^{(0)} + R$. We have already established the general structure of
states on the decoupled level and it is now simple to write down the solution
of the cohomology problem (2.33); it is

$$| \, \alpha \, j \, m >_{\rm ph} \, = \sum_M < j \, M \, j - M \, | \, 00 >
\, | \, \alpha \, j \, M >_{\rm F} \, | \, j - M \, m >_{\rm B} \, | \, 0
>_{\rm gh} \eqno{(2.34)}$$
 with $< j \, M \, j - M \, | \, 00 >$ the
Glebsch--Gordon coefficients.  We note that (2.34) restricts the a priori
infinite number of $SU(2)$ representations carried by boson Fock space to those
carried by fermion Fock space.  Equation (2.34) shows that every state in
Fermion Fock space gives rise to exactly one physical state.  This establishes
the isomorphism between the decoupled formulation and the original model on the
physical Hilbert space.

We note that the BRST condition can also be interpreted as a bosonization rule
which states that on the physical subspace the following replacements may be
made: $j^{(0)}\rightarrow -R$.  This bosonization dictionary can be completed
by constructing physical operators, i.e., the operators that commute with the
BRST charge $Q_1$.  Once this has been done, a set of rules result according to
which every fermion operator can be replaced by an equivalent bosonic operator.
It is easy to check that the following operators are BRST invariant

$$\begin{array}{rcl} 1, & & \eta^\dagger_f \, \eta_f \, ,\\[3mm] \eta_f &=& V
\, \psi_f \, , \, \eta^\dagger_f \; = \; \eta^\dagger_f \, V^{-1} \, ,\\[3mm]
j^a_f &=& \psi^\dagger_f \, V^{-1} \, t^a \, V \, \psi_f \; = \; \eta^\dagger_f
\, t^a \, \eta_f \, ,\\[3mm] L^a &=& \frac{1}{g} \, {\rm tr} \, (V \, i \,
\partial_t \, V^{-1} \, t^a) \, . \end{array} \eqno{(2.35)}$$

We recognize in $\eta_f$, $L^a$ and $j^a_f$ the (physical) fermion fields,
boson fields $B^a = g \, L^a$, and generators of the $SU(2)$ color symmetry
associated with the partition function (2.6) of the original model.  Note
that the currents $R^a = \frac{1}{g} \, {\rm tr} \, (i \, \partial_t \,
V^{-1}) \, V \, t^a$ appearing in the BRST charge are not BRST invariant.  Once
the physical operators have been identified, the physical Hilbert space can be
constructed in terms of them.  In this way the isomorphism (2.34) can also be
established.

As we have now demonstrated explicitly the only condition that physical states
are required to satisfy in order to ensure the above isomorphism is that $Q_1
\, | \, \Psi >_{\rm ph} = 0$.  It is, however, interesting to examine what
further
restrictions would result by imposing that a state be annihilated by all three
nilpotent charges.  Since these charges do not commute, it raises the question
as to whether this is a consistent requirement.  This leads us to consider the
algebra of those BRST charges. One finds

$$\begin{array}{rcl} [Q_{\alpha} \, , \, Q_{\beta}] &=& K_{(\gamma)}
\;\, (\alpha \, , \, \beta \, , \, \gamma \quad \mbox {\rm cyclic}) \, ,\\[3mm]
K_{(\gamma)} &=& - \frac{1}{2} \, f^{abc} \, J^a_{(\gamma)} \, c^b \, c^c
\end{array} \eqno{(2.36)}$$
where $J^a_{(\gamma)} = j^a \, , \, R^a$ and $j^a + R^a$ for $\gamma = 1 \, ,
\, 2 ~\mbox {\rm and} ~3 \, ,$ respectively.  The $K_{(\gamma)}$ are nilpotent
and further have the properties $[K_{(\gamma)} \, , \, Q_{\alpha} ] =
0$ and $[K_{(\gamma)} \, , \, K_{(\gamma^\prime)} ] = 0$.

The $K_{(\gamma)}$ generate the infinitesimal transformation

$$\begin{array}{rclcl} [K_1, \, \psi] &=& -\frac{1}{2} \, \{c \, , \, c \}
\, \psi \, , \quad
[K_1, \, b] &=& \{j \, , \, c\} \, ,\\[5mm]
[K_2, \, \psi]
&=& \frac{1}{2} \, \{c \, , \, c \} \, V \, , \quad
[K_2, \, b] &=& \{R \, , \,
c\} \end{array} \eqno{(2.37)}$$
with $K_1 + K_2 = K_3$. As before $[~,~]$ denotes a graded commutator and
$\{~,~\}$ a matrix anti-commutator. All other transformations vanish. They
are easily checked to represent a symmetry of the action, as is required by
consistency.

From eq (2.37) we note that the conditions $Q_{\alpha} \, | \, \Psi > \,
= 0 \;
(\alpha = 1 \, , \, 2 \, , \, 3)$ can only be consistently imposed if we
require $K_{(\gamma)} \, | \, \Psi > \, = 0 \; (\gamma = 1 \, , \, 2 \, , \,
3)$ as well.  The implementation of all three conditions would
restrict the physical (ghost number zero) states to be
singlets with respect to the physical fermionic currents generating the $SU(2)$
symmetry.
From eq (2.1) we note that for $g > 0$ the ground--state is a singlet.
There is in
fact a double degeneracy since there is a double multiplicity in the singlet
sector for an arbitrary number of flavors.  By restricting to this subspace
one is therefore effectively studying the ground--state sector of the model.

\section{Non-abelian bosonization by coset factorization}

Consider the partition function of free fermions in the fundamental
representation of $U(N)$.
As mentioned in the introduction, the  approach of ref. \cite{10, 12}
 to the bosonization of such fermions in two dimensions
 is most naturally implemented by factoring
 from the corresponding partition function $Z^{(0)}_F$  a topological
 $U(N)/U(N)$ coset carrying the fermion and chiral selection rules
associated with the fermions, but no dynamics \cite{11}.
In factoring out this coset, the bosonization BRST symmetry of ref.
\cite{10,12} is also uncovered.

To emphasize the care with which BRST symmetries have to be implemented
in the identification of
physical states, we note that if we were to ignore the BRST constraint
linking the coset sector to the remaining WZW sector , we would conclude
that the
spectrum of free fermions is not equivalent to a WZW model, but to
the direct product of the coset model and the WZW model. In particular we
would conclude
that this spectrum in N-fold degenerate.
A correct interpretation thus requires a careful analysis of
the BRST symmetries associated
with the introduction of additional degrees of freedom in the path integral,
and those associated with changes of variables.
As we show in this section the original spectrum of free fermions
is obtained, if the BRST symmetries of the physical states
are correctly identified.

To discuss the BRST cohomology associated with
the bosonization BRST, it is useful to decouple the coset again, that is,
we work with the fermionic coset in its decoupled form.
The reason for doing this is that it is
more convenient to analyze the physical spectrum of the coset model
in the decoupled formulation
\cite{9}.

As in the quantum mechanical models discussed above, one can proceed with the
bosonization procedure and only after the final action has been obtained,
the BRST symmetries  are identified by inspection.  The disadvantage of this
procedure, as became abundantly clear in our discussion above, is that one does
not recognize which of these BRST charges should be imposed as symmetries of
the physical states to ensure equivalence with the original free fermion
dynamics.  Instead we follow here the procedure used above to identify
the relevant BRST charges from first principles.

In subsection 3.1 we  review briefly the main results of \cite{11}, showing
how the bosonization BRST arises by an argument similar to that of section 2.
In subsection 3.2 we proceed to rewrite the
coset in the decoupled form, keeping track of the bosonization BRST and the
new BRST that arises when the decoupling is performed. In the last part of
this section we briefly discuss the structure of the physical Hilbert space.

\subsection{Bosonization BRST}

As explained in ref. \cite{11} the partition function of free Dirac
fermions  in the fundamental
representation of $U(N)$ can be written as:
$$
Z_F^{(0)}=Z_{U(N)/U(N)} \times Z_{\rm WZW} \eqno{(3.1.1{\rm a})}
$$
where $Z_{U(N)/U(N)}$ is the partition function of a ${U(N)}/{U(N)}$
coset,
$$\begin{array}{rcl}
Z_{U(N)/U(N)} &=&\int [d\eta][d\bar\eta] \int [d(ghosts)] \int [dB_-]\\
&&\times\, e^{i\int d^2x\{ \eta^\dagger_1 i\partial_+ \eta_1
+\eta^\dagger_2 (i\partial_- + B_-){\eta}_2\}}
e^{i\int d^2x {\rm tr} b_- i\partial_+ c}\end{array}
\eqno{(3.1.1{\rm b})}
$$
and $Z_{WZW}$ is the partition function
$$
Z_{WZW} = \int [dg] e^{i\Gamma[g]} \eqno{(3.1.1{\rm c})}
$$
of a Wess-Zumino-Witten (WZW) field $g$ of level one,
with $\Gamma [g]$ the corresponding action \cite{21}
$$\begin{array}{rcl}
\Gamma[g]& = & \frac{1}{8\pi}\int\!d^2 x\,{\rm tr}(\partial_{\mu}
         g^{-1}\partial^{\mu} g^{-1})\\
    && +\frac{1}{12\pi}\int_{\Gamma}\!d^3 x\, \epsilon^{\mu \nu\rho}
       {\rm tr}(g\partial_{\mu} g^{-1}g\partial_{\nu}g^{-1}
       g\partial_{\rho} g^{-1})  \quad. \end{array}
\eqno{(3.1.2)}
$$
Since, as we have seen in section 2,
the Hilbert space of the bosonic sector
(described by the WZW action in the case in question) is in general
much larger than that of the original fermionic description,
the question arises
as to which constraints must be imposed in order to ensure equivalence
of the two formulations. We now show that if the BRST symmetry of the
physical states is correctly identified, the original spectrum of free
fermions is recovered.

We begin by briefly reviewing the steps leading to the factorized
form (3.1.1), with the objective of establishing systematically which
of the
BRST symmetries should be imposed on the physical states.

We start with the partition function of
free Dirac fermions in the fundamental
representation of $U(N)$,
$$
Z_{\rm F}^{(0)}=\displaystyle\int\,[d\eta][d\bar\eta]
e^{i\int d^2x [\eta^{\dagger}_1 i \partial_- \eta_1
+  \eta^{\dagger}_2 i \partial_+ \eta_2}]\,. \eqno{(3.1.3)}
$$
Following ref. \cite{11} we enlarge
the space by introducing bosonic $U(N)$
Lie algebra valued fields $B_-=B_-^at^a$ (${\rm tr}(t^at^b)=\delta^{ab}$)
via the
identity
$$
1 = \displaystyle \int \, [d \, B_-] \, e^{i\int \, d^2
\, x \, [\eta^\dagger_2 \, B_- \, \eta_2]} \, \delta \, [B_-]\,.\eqno{(3.1.4)}
$$

Using a Fourier representation of the Dirac
Delta functional by introducing an
auxiliary field $\lambda_+$, the partition function
(3.1.3) then takes
the alternative form
$$
Z^{(0)}=\int [d\eta][d\bar\eta]\int [d\lambda_+][dB_-]
e^{i\int\{\eta^\dagger_1i\partial_+\eta_1
+\eta^\dagger_2(i\partial_- +B_-)\eta_2 + {\rm tr} \lambda_+
B_-\}}\eqno{(3.1.5)}
$$
where $\lambda_+$ are again $U(N)$ Lie algebra
valued fields.

We now make the
change of variable $\lambda_+ \to g$ defined by $\lambda_+=\alpha
g^{-1}i\partial_+ g$
where $g$ are $U(N)$ group-valued fields.
The Jacobian associated with this transformation is ambiguous
since we do not have gauge invariance as a guiding principle.
For reasons to become apparent later, we
choose it to be defined with respect
to the Haar measure $g\delta g^{-1}$. Noting that
$$
\delta (g^{-1}i\partial_+ g) = -g^{-1}
i\partial_+(g\delta g^{-1})g\eqno{(3.1.6)}
$$
we have for the corresponding Jacobian,
$$
J=\int[d\tilde b_-][d c_-]
e^{i\int {\rm tr}(g\tilde b_-g^{-1})\partial_+ c_-}\,. \eqno{(3.1.7)}
$$
The partition function (3.1.5) then takes the form
$$\begin{array}{rcl}
Z^{(0)}_F&=&\int [d\eta][d\bar\eta]\int [d(ghosts)]\int [dg][dB_-]
e^{i\int d^2x\{\eta^\dagger_1i\partial_+\eta_1
+ \eta^\dagger_2(i\partial_- +B_-)\eta_2\}}\\
&&\times\, e^{i\int d^2x\{\alpha {\rm tr} (g^{-1}i\partial_+gB_-)+
{\rm tr}\tilde b_-g^{-1}i(\partial_+c_-)g\} }\,.\end{array}
\eqno{(3.1.8)}
$$
There is a BRST symmetry associated with the change of variable $\lambda_+
\to g$.
In order to discover it we systematically
perform this change of variable
by introducing  in (3.1.5) the identity
$$
1=\int[dg]J\delta[\lambda_+ -\alpha g^{-1}i\partial_+g] \eqno{(3.1.9)}
$$
where $J$ is the Jacobian defined in (3.1.7).
Using the Fourier representation
for the delta functional we are thus led to the alternative form for the
partition function,
$$\begin {array}{rcl}
Z^{(0)}&=&\int [d\eta][d\bar\eta]\int [d(ghosts)]\int [dg][dB_-]\int
[d\lambda_+][d\rho_-]
 e^{iS_{{\rm aux}}}\\
&&\times\,e^{i\int\{\eta^\dagger_1i\partial_+\eta_1
+\eta^\dagger_2(i\partial_- +B_-)
\eta_2+{\rm tr} \lambda_+B_-\}}\end {array}
\eqno{(3.1.10)}
$$
where
$$
S_{{\rm aux}}=\int d^2x {\rm tr}\{\rho_-(\lambda_+ - \alpha
g^{-1}i\partial_+g) +
 g\tilde b_-g^{-1}i\partial_+c_- \}\,.\eqno{(3.1.11)}
$$
The  auxiliary action, $S_{{\rm aux}}$, is
evidently invariant under the off-shell
nilpotent transformations
$$\begin{array}{rcl}
&&\delta_1B_-=\delta_1\rho_-=\delta_1\lambda_+ =\delta_1\eta_1 =
\delta_1\eta_2 = 0\,,\\
&&\delta_1 g g^{-1} = c_-\,,\\
&&\delta_1 \tilde b_- = \alpha \rho_-,
\quad \delta_1 c_- = {1\over 2}{\{c_- , c_-\}}\,.
\end{array}\eqno{(3.1.12)}
$$
We now observe that $S_{{\rm aux}}$ may be written as
$$
S_{{\rm aux}} = {1\over\alpha}
{\delta_1{\rm tr}\tilde b_- (\lambda_+ - \alpha g^{-1}i\partial_+ g)}\,.
\eqno{(3.1.13)}
$$
Hence $S_{{\rm aux}}$ is BRST exact, so that equivalence of the two partition
functions is guaranteed on the (physical) states invariant under the
transformations (3.1.12).

Integrating over $\rho_-$ and $\lambda_+$
the BRST transformations (3.1.12) are replaced by
$$\begin{array}{rcl}
&&\delta_1 B_- = \delta_1\eta_1 = \delta_1\eta_2 = 0\,,\\
&&\delta_1g g^{-1} = c_-\,,\\
&&\delta_1 \tilde b_- = -\alpha B_-,
\quad \delta_1 c_- = {1\over 2}{\{c_- , c_-\}}\,.\end{array}
\eqno{(3.1.14)}
$$
and the partition function (3.1.10) reduces to
(3.1.8).
We now further make the change of variables
$$
\eta_2 \to {\eta'}_2 = g\eta_2,
\quad \tilde b_- \to b_- = g\tilde b_- g^{-1}\,.\eqno{(3.1.15)}
$$
The transformation $\tilde b_- \to b_-$ has
Jacobian one. The Jacobian
associated with the transformation
$\eta_2 \to {\eta'}_2$ is, on the other hand, given by
$$
J_F = e^{i\Gamma[g]-\frac
{i}{4\pi}\int d^2x {\rm tr}(B_- g^{-1}i\partial_+ g)}\,. \eqno{(3.1.16)}
$$
Notice that $J_F$ contains the contribution
from the non-abelian, as well as
abelian $U(1)$ anomaly. For the choice $\alpha = {1\over {4\pi}}$ the
second term in $\ln J_F$ cancels the term proportional to $\alpha$ in
(3.1.8).
Noting that $g(i\partial_- + B_-)g^{-1} = i\partial_- + B_-'$ with
$B_-' = gB_-g^{-1} + gi\partial_- g^{-1}$, using $[dB] = [dB']$, and
streamlining the notation by dropping "primes" everywhere,
 the partition function (3.1.8)
 reduces to (3.1.1a), and
 the BRST transformations (3.1.14) now read in terms of the new variables,
$$\begin{array}{rcl}
&&\delta_1g g^{-1} = c_-\,,\\
&&\delta_1 \eta_2 = c_- \eta_2, \quad \delta_1 \eta_1 = 0\,,\\
&&\delta_1 b_- = -\frac{1}{4\pi}B_- +
\frac{1}{4\pi}gi\partial_- g^{-1} + \{b_-, c_-\}\,,\\
&&\delta_1 c_- = \frac{1}{2}\{c_-, c_-\}\,,\\
&&\delta_1 B_-= [c_-, B_-] - i\partial_- c_-\,.
\end{array}\eqno{(3.1.17)}
$$
As one readily checks, they represent a symmetry
of the partition function (3.1.1a).  We see that these BRST conditions
couple the matter sector ($g$) to the coset sector.
As we have shown, they must be symmetries of the physical states.

\subsection{BRST analysis of coset sector}

It is inconvenient to analyze the cohomology problem with the $U(N)/U(N)$
coset realized in the present form as a constrained
fermion system. Instead it is
preferable to decouple \cite{9} in (3.1.1b) the $B_-$
field from the fermions, in order
to rewrite the coset partition function in
terms of free fermions, negative
level WZW fields and ghosts. As we now show, this procedure
leads to an additional BRST symmetry.

Concentrating now on the coset sector we
introduce in (3.1.1b) the identity
$$
1=\int [d \rho_+][d h] [d \tilde b_+] [d \tilde c_+]
e^{i\tilde S_{\rm aux}}\eqno{(3.2.1)}
$$
with
$$
\tilde S_{\rm aux}= \displaystyle\int\,d^2x {\rm tr}(\rho_+
[B_- -  h i\partial_-h^{-1}])
+ {\rm tr}(\tilde b_+ iD_-(h) \tilde c_+)\eqno{(3.2.2)}
$$
where $h$ is a $U(N)$ group-valued field,
$D_-(h)=\partial_-+[h\partial_- h^{-1},]$ and, like the $\tilde
b_-,c_-$--ghosts,
the $\tilde b_+,\tilde c_+$--ghosts
transform in the adjoint representation.
Note that, unlike in the previous case,
the representation of the coset
as a gauged fermionic system has led us
to define the Jacobian with respect
to the Haar measure $h\delta h^{-1}$:
$$
\delta(h\partial_- h^{-1}) = \det D_-(h)h\delta h^{-1} \,.\eqno{(3.2.3)}
$$
 $\tilde S_{{\rm aux}}$ is invariant under the off-shell nilpotent
transformation
$$\begin{array}{rcl}
\delta_2 \rho_+ &=& \delta_2 B_- = 0 \,,\\
h\delta_2 h^{-1} &=& \tilde c_+ \,,
\quad
\delta_2 \tilde b_+ =\rho_+\,,
\quad
\delta_2 \tilde c_+ = {-\frac{1}{2}}{\{\tilde c_+,\tilde c_+\}}
\end{array}\eqno{(3.2.4)}
$$
and $\tilde S_{{\rm aux}}$ is readily seen to be an exact form with respect
to this transformation. As before we thus conclude that the
original dynamics is recovered on the subspace
annihilated by the corresponding
BRST charge.

Following the previous steps we find, upon
introducing the identity (3.2.1)
in (3.1.1b) and integrating over $\rho_+$ and $B_-$,
$$\begin{array}{rcl}
Z_{U(N)/U(N)} &=& \int [d\eta][d\bar\eta]\int [dh] [d(ghosts)]
e^{i\int d^2x\{\eta_1^\dagger i\partial_+ \eta_1 +
{\eta}_2^\dagger (i\partial_- + hi\partial_- h^{-1}){\eta}_2}\\
&&\times\,e^{i\int d^2x \{{\rm tr}(b_-i\partial_+ c_-) + {\rm tr}(\tilde
b_+ iD_-(h)
\tilde c_-)\}}\,.
\end{array}\eqno{(3.2.5)}
$$
This partition function is seen to be invariant
under the BRST transformation
$$
h\delta_2 h^{-1} = \tilde c_+\,,
\quad
\delta_2 \tilde b_+ = {\eta}_2 {\eta}_2^\dagger \,,
\quad
\delta_2 \tilde c_+ = {-\frac{1}{2}}{\{\tilde c_+,\tilde c_+\}} \,,
\delta_2 \eta_1 = 0, \quad \delta_2 {\eta}_2 = 0\eqno{(3.2.6)}
$$
obtained from (3.2.4) after making use of the
 equations of motion associated with a general variation in $\rho_+$ and $B_-$.

We now decouple the fermions by making the change of variable
${\eta}_2 \to \psi_2$ with
${\eta}_2 = h\psi_2$.
Correspondingly the last variation in (3.2.6) is replaced by
$\delta_2 \psi_2 = -h^{-1} \tilde c_+ h$. Taking account of the Jacobian
exp${-i\Gamma[h]}$ associated with this change of variable
and setting $\eta_1 = \psi_1$ to further streamline the notation,
the coset partition function (3.2.5) reduces to
$$\begin{array}{rcl}
Z_{U(N)/U(N)} &=&\int[d\eta][d\bar\eta] \int [d(ghosts) \int [dh]
e^{-i\Gamma[h]}\\
&&\times\,e^{\int d^2x
\{\psi_1^\dagger i\partial_+ \psi_1 + \psi_2^\dagger i\partial_- \psi_2
+ {\rm tr}(b_- i\partial_+ c_-) + {\rm tr}(\tilde b_+ D_-(h) \tilde c_+)\}
}\,.  \end{array}\eqno{(3.2.7)} $$ Finally we also decouple the ghosts $\tilde
b_+, \tilde c_+$ by making the change of variables $\tilde b_+ \to b_+, \tilde
c_+ \to c_+$ defined by $$ \tilde b_+ = hb_+ h^{-1} , \quad \tilde c_+ = hc_+
h^{-1}\,.\eqno{(3.2.8)} $$ Only the $SU(N)$ part of $h$ contributes a
non-trivial Jacobian. Setting $h = v\hat h$ with $v \,\in\, U(N)$ and $ \hat
h\, \in\, SU(N)$, we have $$ [d\tilde b_+][d\tilde c_+] = e^{-iC_V \Gamma[\hat
h]} [d b_+][d c_+]\eqno{(3.2.9)} $$ where $C_V$ is the quadratic Casimir in the
adjoint representation.  Making further use of the Polyakov-Wiegmann identity
one has $\Gamma[h] = \Gamma[\hat h] + \Gamma[v]$ and our final result for the
coset partition function reads $$ Z_{U(N)/U(N)} = \int [d\eta][d\bar\eta]\int
[d(ghosts)]\int [dv][d\hat h] e^{iS_{U(N)/U(N)}} \eqno{(3.2.10)} $$ with
$$\begin{array}{rcl} S_{U(N)/U(N)} &=& -\Gamma[v]-(1+C_V)\Gamma[\hat h] + \int
d^2x \{ \psi_1^\dagger i\partial_- \psi_1 + \psi_2^\dagger i\partial_+
\psi_2\}\\ &+&\int d^2x\{ {\rm tr}(b_- i\partial_+ c_-) + {\rm tr}(b_+
i\partial_- c_+) \}\,.\end{array}\eqno{(3.2.11)} $$ Notice that $v$ and  $\hat
h$ correspond to level -1 and -(1+$C_V$) fields, respectively. Notice also that
the ghost term contains the $SU(N)$ as well as $U(1)$ contributions.

In terms of the new variables the BRST conditions (3.2.6) read (notice
in particular the changes with regard to the first and last variations)
$$\begin{array}{rcl}
&&h^{-1} \delta_2 h = -c_+ \,,\\
&&\delta_2\psi_1 = 0 \,,
\quad
\delta_2\psi_2 = c_+\psi_2 \,,\\
&&\delta_2 b_+ = \psi_2\psi_2^\dagger
- \frac{1}{4\pi}v^{-1} i\partial_+ v
- \frac{(1+C_V)}{4\pi}\hat h^{-1} i\partial_+ \hat h
+ \{b_+,c _+\} \,,\\
&&\delta_2 c_+ = \frac{1}{2}\{c_+,c_+\} \,.\\
\end{array}\eqno{(3.2.12)}
$$
Notice that we have included in the
transformation law for $\delta_2 b_+$ an
anomalous piece proportional to $(1+C_V)$, in order to
compensate the contribution
coming from the variation of the (anomalous) first two terms in (3.2.11)
arising from the Jacobians of the transformations.

Finally, returning to the transformation laws (3.1.17), and
recalling that, according to (3.2.2)
$B_- = hi\partial_- h^{-1}$, these transformations
are to be replaced by
$$\begin{array}{rcl}
&&g\delta_1 g^{-1} = -c_- ,\quad
h\delta_1 h^{-1} = -c_- \,,\\
&&\delta_1\psi_1 = \delta_1\psi_2 = 0\,,\\
&&\delta_2b_- = \frac{1}{4\pi} gi\partial_- g^{-1}
 -\frac{(1+C_V)}{4\pi}hi\partial_- h^{-1}
+ \{b_-,c_-\} \,,\\
&&\delta_1c_- = \frac{1}{2}\{c_-,c_-\}\,.
\end{array}\eqno{(3.2.13)}
$$
where an anomalous piece has again been included in the variation for $b_-$
in order to compensate for the corresponding contribution coming from the
Jacobian in (3.2.9).

The corresponding BRST charges are obtained
via the usual Noether construction,
and are found to be of the general form
$$
\Omega_{\pm} = {\rm tr} c_{\pm} [\Omega_{\pm} -
\frac{1}{2}\{c_{\pm},c_{\pm}\}]\eqno{(3.2.14)}
$$
for the
$SU(N)$ and $U(1)$ pieces separately. For the $U(1)$
piece the anticommutator of
the ghosts vanishes, of course. Setting
$g = u\hat g$, $u\, \in\, U(1)$ ,
$\hat g \,\in\, SU(N)$ and noting that $\Gamma[g] = \Gamma[u] +
\Gamma[\hat g]$, we find
$$\begin{array}{rcl}
&&\Omega_- = \frac{1}{4\pi}ui\partial_- u^{-1} - \frac{1}{4\pi}vi\partial_-
v^{-1}\,,\\
&&\Omega_+ = {\rm tr}(\psi_2 \psi_2^\dagger) -
\frac{1}{4\pi}v^{-1}i\partial_+ v\,,\\
&&\Omega_-^a = {\rm tr} t^a [\frac{1}{4\pi}gi\partial_- g^{-1}
- \frac{(1+C_V)}{4\pi} hi\partial_- h^{-1} + \{b_-,c_-\}]\,,\\
&&\Omega_+^a = {\rm tr} t^a[\psi_2\psi_2^\dagger -
\frac{(1+C_V)}{4\pi}h^{-1}i\partial_+h
+ \{b_+,c_+\}]\end{array}\eqno{(3.2.15)}
$$
where $a = 1,...,N^2-1$ . All these operators are required to annihilate
the physical states, as we have seen.
By going over to canonical variables
and using the results of ref. \cite{9},
one easily verifies that these constraints
are first class with respect to themselves (vanishing central extension).
Indeed, define (tilde stands for ``transpose'')
$$\begin{array}{rcl}
&&\tilde{\hat\Pi}^{g}=\frac{1}{4\pi}\partial_0 g^{-1},\,,\\
&&\tilde{\hat\Pi}^{h}=-\frac{(1+C_V)}{4\pi}\partial_0
h^{-1}\,.\end{array}\eqno{(3.2.16)}
$$
Canonical quantization then implies the Poisson algebra (see ref.
\cite{2,22} for derivation; $g$ stands for a generic field)
$$\begin{array}{rcl}
&&\{g_{ij}(x),\hat\Pi^{g}_{kl}(y)\}_P=\delta_{ik}\delta
_{jl}\delta(x^1-y^1)\,,\\
&&\{\hat\Pi_{ij}^{g}(x),\ \hat\Pi^{g}_{kl}(y)\}_P
=-\frac{1}{4\pi}\left(\partial_1g^{-1}_{jk}g^{-1}_{li}-g_{jk}
^{-1}\partial_1g^{-1}_{li}\right)\delta (x^1-y^1)\,. \end{array}\eqno{(3.2.17)}
$$
In terms of canonical variables, we have for the constraints (3.2.15)
$$\begin{array}{rcl}
&&\Omega_+= {\rm tr}[\psi_2\psi^\dagger_2
- i\tilde{\hat\Pi}^{h}h -\frac{1}{4\pi}h^{-1}i\partial_1 h]\,,\\
&&\Omega_-= {\rm tr}[ig\tilde\Pi^{g} - \frac{1}{4\pi} gi\partial_1 g^{-1}
+ih\tilde{\hat\Pi}^{h}+\frac{1}{4\pi}hi\partial_1 h^{-1}]\\
&&\Omega^a_+= {\rm tr}t^a[\psi_2\psi^\dagger_2
- i\tilde{\hat\Pi}^{h}h -\frac{(1+C_V)}{4\pi}h^{-1}i\partial_1 h
+ \{b_+,c_+\}]\,,\\
&&\Omega^a_-= {\rm tr}t^a[ig\tilde\Pi^{g} - \frac{1}{4\pi} gi\partial_1 g^{-1}
+ih\tilde{\hat\Pi}^{h}+\frac{(1+C_V)}{4\pi}hi\partial_1 h^{-1}
+ \{b_-,c_-\}]\,.\end{array}\eqno{(3.2.18)}
$$
With the aid of the Poisson brackets (3.2.17) it is straightforward
to verify that $\Omega_{\pm}$ and $\Omega^a_{\pm}$ are
fist class:
$$
\left\{\Omega_\pm(x),\Omega_\pm(y)\right\}_P=0, \quad
\left\{\Omega_\pm^a(x),\Omega^b_\pm(y)\right\}_P
=-f_{abc}\Omega^c_\pm\delta(x^1-y^1) \,.\eqno{(3.2.19)}
$$
 Hence the corresponding BRST charges are nilpotent.

The physical Hilbert space is now obtained by solving the cohomology problem
associated with the BRST charges $\Omega_\pm$ in the
ghost-number zero sector.   This can be done in two ways.  One can either
 solve the cohomology problem
in the Hilbert space explicitly, i.e., find the states which are annihilated
 by the BRST charges, but which
are not exact, as was done in section 2.  Alternatively one can construct
the physical operators, which
commute with the BRST charges, in terms of which the physical Hilbert space can
 be constructed.
Here we prefer  to follow the second approach as it is more transparent.
For completeness let us indicate
how  the analysis would proceed in the first approach.

Although technically more involved, the analysis parallels that of section 2.
 One first notes  that on the
decoupled level the Hilbert space is the direct product of four sectors,
namely, a free fermion sector,
positive - and negative level WZW sectors and a ghost sector.  Each of the
sectors
is again the direct
product of  left and right moving sectors.  For the matter fields the left
and right
 moving sectors are not
independent, but have to be combined in a specific way, namely,
they must belong to the same
representation of the Kac-Moody algebra \cite{23}.  This is analogous
to the quantum mechanical model
discussed in section 2 where the left and right symmetries also belong to
the same $SU(2)$ representation.
Using this and the results of ref. \cite{9} one finds that the constraints
(3.2.18) relate the representations
and weights of the various sectors in such a way that a one-to-one
correspondence
is established between
the states of the free fermion model and the physical states annihilated by
the BRST charges (3.2.18).

This equivalence is even more transparent  in the second approach where one
 requires that the
physical operators commute with the constraints.
Making use of the Poisson brackets (3.2.17), we have
$$\begin{array}{rcl}
&&\left\{\Omega_-^a(x),\ g^{-1}(y)\right\}_P=
i(g^{-1}(x)t^a)\delta(x^1-y^1)\,,\\
&&\left\{\Omega_-^a(x),\ h(y)\right\}_P=
-i(t^a h(y))\delta(x^1-y^1)\end{array}\eqno{(3.2.20)}
$$
$$\begin{array}{rcl}
&&\left\{\Omega_+^a(x),\ \psi(y)\right\}_P=
-i(t^a\psi)\delta(x^1-y^1)\,,\\
&&\left\{\Omega_+^a(x),\ h(y)\right\}_P=
i(h(y)t^a)\delta(x^1-y^1)\end{array}\eqno{(3.2.21)}
$$

From the Poisson brackets (3.2.20) follows  that the fields $h$ and $g$
can occur in physical observables only in the combinations $g^{-1}h$.
From the other two Poisson brackets (3.2.21) follows
that the fermion field can
only occur in the combination $h\psi$. Putting things together we conclude
that physical fermion field corresponds to the local product $g^{-1}h\psi$.
Turning back our set of transformations on the original fermion field,
we see that the BRST conditions establish in this way a one-to-one
correspondence between the fields of the decoupled formulation and the
original free fermion field: $\eta = h^{-1}g\psi$.
This establishes the equivalence of the decoupled partition function
(3.1.1a), subject to the BRST conditions, to the fermionic one as given by
(3.1.3). The coset factor in (3.1.1a) merely encodes the selection rules
of the partition function (3.1.3), but carries no dynamics.

\section{$QCD_2$ in the local decoupled formulation}

As a final example we prove deductively that the BRST charges
associated with the currents (2.40) and (2.49) of ref. \cite{13}
must annihilate the physical states in order to ensure
equivalence with the original formulation. To this end
we start from the partition $QCD_2$ function
$$
Z=\int [dA_+][dA_-]\int[d\psi][d\bar\psi] e^{iS[A,\psi,\bar
\psi]}\eqno{(4.1)}
$$
with
$$
S[A,\psi,\bar\psi]=-\frac{1}{4}{\rm tr} F_{\mu\nu}F^{\mu\nu}
+\psi^\dagger_1(i\partial_++eA_+)\psi_1+\psi^\dagger
_2(i\partial_-
+eA_-),\eqno{(4.2)}
$$
where $F_{\mu\nu}$ is the chromoelectric field strength
tensor, and $\partial_\pm=\partial_0\pm
\partial_1,\\
 A_\pm=A_0\pm A_1$.

We parametrize $A_\pm$ as follows:
$$
eA_+=U^{-1}i\partial_+U,\quad eA_-=Vi\partial_-V^{-1}\eqno{(4.3)}
$$
and change variables from $A_\pm$ to $U,V$
by introducing the identities
$$
\begin{array}{rcl}
&&1=\int[dU]\det iD_+(U)\delta(eA_+
-U^{-1}i\partial_+U)\,,\\
&&1=\int[dV]\det iD_-(V)\delta(eA_-
-Vi\partial_-V^{-1})\end{array}\eqno{(4.4)}
$$
in the partition function (4.1). Here $D_+(U)$ and
$D_-(V)$ are the covariant derivatives in the adjoint
representation:
$$\begin{array}{rcl}
&&D_+(U)=\partial_++[U^{-1}\partial_+U,\,],\\
&&D_-(V)=\partial_-+[V\partial_-V^{-1},\,].\end{array}\eqno{(4.5)}
$$
Exponentiating as usual the corresponding functional
determinants in terms of ghost fields and
representing the delta functions as a Fourier integral, we
obtain for (4.1)
$$\begin{array}{rcl}
Z&=&\int[dA_+][dA_-][d\psi][d\bar\psi]\int[d
U][dV][d\lambda_+][d\lambda_-]\int[d(ghosts)]\\
 &&\times e^{iS[A,\psi,\bar\psi]}
\times e^{i\int\lambda_-(eA_+-U^{-1}i\partial_+U)+i\int b_-iD_+
(U)c_-}\\
&&\times e^{i\int\lambda_+(eA_--Vi\partial_-V^{-1})+i\int b_+iD_-
(V)c_+},\end{array}\eqno{(4.6)}
$$

We follow again the procedure of ref. \cite{24}. The partition
function (4.6) is seen to be invariant under the
transformations
$$\begin{array}{rcl}
&&\delta_1\lambda_+=0,\ \delta_1A_-=0\,,\\
&&V\delta_1V^{-1}=c_+\,,\nonumber\\
&&\delta_1\psi_2=0\,,\\
&&\delta_1b_+=\lambda_+\,,\nonumber\\
&&\delta_1c_+=-\frac{1}{2}\{c_+,c_+\}\end{array}\eqno{(4.7{\rm a})}
$$
and
$$\begin{array}{rcl}
&&\delta_2\lambda_-=0,\ \delta A_+=0\,,\nonumber\\
&&U^{-1}\delta_2U=c_-\,,\nonumber\\
&&\delta_2\psi_1=0\,,\\
&&\delta_2b_-=\lambda_-\,,\nonumber\\
&&\delta_2c_-=-\frac{1}{2}\{c_-,c_-\}\,.\end{array}\eqno{(4.7{\rm b})}
$$
These transformations are off-shell nilpotent. It
is easily seen that in terms of the graded variational
derivatives $\delta_{1,2}$, the effective action in
(4.6) can be rewritten as
$$
S_{eff}=S[A,\psi,\bar\psi]+\Delta_1+\Delta_2\eqno{(4.8)}
$$
where
$$\begin{array}{rcl}
&&\Delta_1=\delta_1[b_-(eA_+-U^{-1}i\partial_+U)]\,,\\
&&\Delta_2=\delta_2[b_+(eA_--Vi\partial_-V^{-1})]\end{array}\eqno{(4.9)}
$$
are $Q_1$ and $Q_2$ exact with  $Q_1,Q_2$ the
BRST Noether charges associated with the respective
transformations. Hence the physical states must belong
to ${\rm kern}\ Q_1/{\rm Im}\ Q_1$ and ${\rm kern}\ Q_2/{\rm Im}\ Q_2$ if
$S_{eff}$ is to be equivalent to the original action
$S[A,\psi,\bar\psi]$.

Integrating over $A_\pm$ and $\lambda_\pm$ the partition
function and BRST transformations reduce to
$$\begin{array}{rcl}
Z&=&\int [dU][dV]\int[d\psi][d\bar\psi]\int
[d(ghosts)]\\
&&\times e^{\frac{i}{2}\int {\rm tr}(F_{01})^2}
e^{i\int(U\psi_1)^\dagger
i\partial_+(U\psi_1)+i\int(V^{-1}\psi_2)^\dagger i\partial_+
(V^{-1}\psi_2)}\\
&&\times e^{i\int b_-iD_+(U)c_-}e^{i\int b_+iD_-(V)c_-}\end{array}\eqno{(4.10)}
$$
and
$$\begin{array}{rcl}
&&V\delta_1V^{-1}=c_+\,,\\
&&\delta_1\psi_2=0\,,\\
&&\delta_1b_+=-\frac{1}{2}D_+(U) F_{01}+
\psi_2\psi_2^\dagger\,,\\
&&\delta_1c_+=-\frac{1}{2}\{c_+,c_+\}\,,\end{array}\eqno{(4.11{\rm a})}
$$
$$
\begin{array}{rcl}
&&U^{-1}\delta_2U=c_-\,,\\
&&\delta_2\psi_1=0\,,\\
&&\delta_2b_-=\frac{1}{2}D_-(V) F_{01}+
\psi_1\psi_1^\dagger\,,\\
&&\delta_1c_-=-\frac{1}{2}\{c_-,c_-
\}\,,\end{array}\eqno{(4.11{\rm b})}
$$
respectively. As one readily checks, the partition
function (4.10) is invariant under these
(nilpotent) transformations which, as we have
seen, must also leave ${\Ha}_{phys}$ invariant.

We now decouple the fermions and ghosts by defining
$$\begin{array}{lcr}
&&\psi_1^{(0)}\equiv U\psi_1,\quad \psi_2^
{(0)}=V^{-1}\psi_2\,,\\
&&b_-^{(0)}=Ub_-U^{-1},\quad c_-^{(0)}=Uc_-U^{-1}\,,\\
&&b_+^{(0)}=V^{-1}c_+V,\quad c_+^{(0)}=V^{-1}c_+V\,.\end{array}
\eqno{(4.12)}
$$
Making a corresponding transformation in the measure, we
have
$$
\begin{array}{rcl}
[d\psi_1][d\psi_2]&=&e^{-i\Gamma[UV]}[d\psi_1^{(0)}][d
\psi_2^{(0)}]\\[1mm]
[d(ghosts)]&=&e^{-iC_V\Gamma[UV]}[d(ghosts^{(0)})]\end{array}\eqno{(4.13)}
$$
where $\Gamma[g]$ is the Wess-Zumino-Witten (WZW) functional (3.1.2).
We thus arrive at the decoupled partition function \cite{6,7,13}
$$
Z=Z_F^{(0)}Z_{gh}^{(0)}Z_{U,V}\eqno{(4.14{\rm a})}
$$
where
$$
Z_F^{(0)}=\int[d\psi^{(0)}][d\bar\psi^{(0)}]e^{i\int\bar
\psi i\slp\psi},\eqno{(4.14{\rm b})}
$$
$$
Z_{gh}^{(0)}=\int[d(ghosts)^{(0)}]e^{i\int b_+^{(0)}i\partial_-
c_+^{(0)}}e^{i\int b_-^{(0)}i\partial_+c_-^{(0)}}\eqno{(4.14{\rm c})}
$$
and
$$
Z_{U,V}=\int[dU][dV]e^{-i(1+C_V)\Gamma[UV]}
e^{\frac{i}{2}\int {\rm tr}(F_{01})^2}\,.\eqno{(4.14{\rm d})}
$$
We rewrite $F_{01}$ in terms of $U$ and $V$ by noting that
$$
\begin{array}{rcl}
F_{01}&=&-\frac{1}{2}[D_+(U)Vi\partial_- V^{-1}
-\partial_-(U^{-1}i\partial_+U)]\\
&=&\frac{1}{2}[D_-(V)U^{-1}i\partial_+U-\partial_+(Vi\partial
_-V^{-1})]\end{array}\eqno{(4.15)}
$$
and making use of the identities
$$
\begin{array}{rcl}
&&D_-(V)B=V[\partial_-(V^{-1}BV)]V^{-1}\\
&&D_+(U)B=U^{-1}[\partial_+(UBU^{-1})]U\end{array}\eqno{(4.16)}
$$
as well as
$$
U^{-1}[\partial_+(U\partial_-U^{-1})]U=-\partial_-(U^{-1}\partial
_+U).\eqno{(4.17)}
$$
We thus obtain the alternative expressions
$$
\begin{array}{rcl}
F_{01}&=&-\frac{1}{2}U^{-1}[\partial_+(\Sigma\partial_-\Sigma^{-1})]
U\\
&=&\frac{1}{2}V[\partial_-(\Sigma^{-1}\partial_+\Sigma)]V\end{array}\eqno{(4
.18)}
$$
where $\Sigma$ is the gauge invariant quantity
$\Sigma=UV$.
The term $-(1+C_V)\Gamma[UV]$ in the effective action arising
from the change of variables is of quantum origin and must be
explicitly taken into account when rewriting the BRST transformations
laws (4.11a), (4.11b) in terms of the decoupled
variables. Its contribution to these transformations is obtained
by noting that
$(\delta=\delta_1+\delta_2)$
$$
\begin{array}{rcl}
-(1+C_V)\delta\Gamma[UV]&=&\frac{1+C_V}{4\pi}
\int {\rm tr}\left\{\left[(UV)^{-1}i\partial_+(UV)
\right]i\partial_-c_+^{(0)}\right.\\
&+&\left.\left[(UV)i\partial_-(UV)^{-1}\right]
i\partial_+c_-^{(0)}\right\}.\end{array}\eqno{(4.19)}
$$
We thus find, making use of (4.18) and the
identities (4.16), (4.17),
$$
\begin{array}{rcl}
\delta V^{-1}V&=&c_+^{(0)}\,,\\
\delta\psi_1^{(0)}&=&c_+^{(0)}\psi_2^{(0)}\,,\\
\delta b^{(0)}_+&=&-\frac{1}{2}\Sigma^{-1}
\left[\partial^2_+(\Sigma i\partial_-\Sigma^{-1})\right]
\Sigma-\left(\frac{1+C_V}{4\pi}\right)
\Sigma^{-1}i\partial_+\Sigma\\
&&+\psi_2^{(0)}\psi_2^{(0)+}+\left\{b_+^{(0)},c_+^{(0)}
\right\}\,,\\
\delta c_+^{(0)}&=&\frac{1}{2}\left\{c_+^{(0)},c_+^{(0)}\right\}\,
\end{array}\eqno{(4.20{\rm a})}
$$
$$
\begin{array}{rcl}
\delta U^{-1}U&=&c_-^{(0)}\,,\\
\delta\psi_1^{(0)}&=&c_-^{(0)}\psi_1^{(0)}\,,\\
\delta b^{(0)}_-&=&-\frac{1}{2}\Sigma
\left[\partial^2_-(\Sigma^{-1} i\partial_+\Sigma)\right]
\Sigma^{-1}-\left(\frac{1+C_V}{4\pi}\right)
\Sigma i\partial_-\Sigma^{-1}\\
&&+\psi_1^{(0)}\psi_1^{(0)\dagger}+\left\{b_-^{(0)},c_-^{(0)}
\right\}\,,\\
\delta c_-^{(0)}&=&\frac{1}{2}\left\{c_-^{(0)},c_-^{(0)}\right\}.
\end{array}\eqno{(4.20{\rm b})}
$$
Notice the change in the transformation law for $V$ and $U$,
as well as the change in sign in the
transformation of $c_\pm^{(0)}$.

We now perform a gauge transformation $U\to UG^{-1}$,
$V\to GV$, taking us to the gauge $U=1$ $(G=U)$:
$U\to 1,\qquad V\to \Sigma$.
The decoupled fields evidently remain unaffected by this gauge
transformation. The transformation laws for $V$ and $U$
above are replaced by a single transformation law
$$
\delta \Sigma^{-1}\Sigma=c_+^{(0)}-\Sigma^{-1}c_-^{(0)}\Sigma.\eqno{(4.21)}
$$
Making once more use of the identities (4.16) and (4.17),
we finally obtain for the BRST transformations for the decoupled
fields in the $U=1$ gauge ($c_+^{(0)}$ and $c_-^{(0)}$ are
to be regarded as independent ``parameters''):
$$
\begin{array}{rcl}
&&\delta \Sigma^{-1}\Sigma=c_+^{(0)}\,,\\
&&\delta\psi_1^{(0)}=c_+^{(0)}\psi_2^{(0)}\,,\\
&&\delta b^{(0)}_+=-\frac{1}{2}\Sigma^{-1}
\left[\partial^2_+(\Sigma i\partial_-\Sigma^{-1})\right]
\Sigma+\psi_2^{(0)}\psi_2^{(0)+}+\left\{
b_+^{(0)},c_+^{(0)}\right\}\,,\\
&&\delta c_+^{(0)}=\frac{1}{2}\left\{c_+^{(0)},c_+^{(0)}\right\}.
\end{array}\eqno{(4.22{\rm a})}
$$
$$
\begin{array}{rcl}
&&\Sigma\delta \Sigma{-1}=-c_-^{(0)}\,,\\
&&\delta\psi_2^{(0)}=c_-^{(0)}\psi_2^{(0)}\,,\\
&&\delta b^{(0)}_-=-\frac{1}{2}\Sigma^{-1}
\left[\partial^2_-(\Sigma^{-1} i\partial_+\Sigma)\right]
\Sigma^{-1}+\psi_1^{(0)}\psi_1^{(0)+}+\left\{b_-^{(0)},
c_-^{(0)}\right\}\,,\\
&&\delta c_+^{(0)}=\frac{1}{2}\left\{c_-^{(0)},c_-^{(0)}\right\}.
\end{array}\eqno{(4.22 {\rm b})}
$$
Using again the identities (4.16) and (4.17), we have
$$
\Sigma[\partial_-^2(\Sigma^{-1}\partial_+\Sigma)]\Sigma^{-1}
=D_-(\Sigma)(\partial_+(\Sigma\partial_-\Sigma^{-1})).\eqno{(4.23)}
$$
Comparing our results with those of ref. \cite{13}, we
see that we have recovered the BRST conditions of the local
formulation (eqs. (2.27) and (2.41) of
ref. \cite{13} after identification of $V$ with $\Sigma$ in the
$U=1$ gauge). This establishes that the transformations
(4.22a) and (4.22b) indeed have to be a symmetry
of the physical states, as has been taken for granted
in ref. \cite{13}.

\section{Conclusion}

Much interest has been devoted recently to gauged WZW theories and $QCD_2$ in a
formulation in which various sectors of the theory appear decoupled on the level
of the partition function, and are only linked via BRST conditions associated
with nilpotent charges. In particular, in the case of $QCD_2$ one is thus
led via
the Noether construction to several such conserved charges; however not all them
are required to vanish on the physical subspace. In order to gain further
insight
 into the question as to which BRST conditions must actually be imposed in
order
to ensure equivalence of the decoupled formulation to the original coupled one,
we have examined this question in the context of simple fermionic models.  We
have in
particular exhibited a general procedure for deciding which of the BRST
conditions
are to be imposed, and have thereby shown that this selects in general a subset
of nilpotent charges. By solving the corresponding cohomology problem we have
shown that one recovers the Hilbert space structure of the original models.

We have further demonstrated  that the requirement that all
of the nilpotent charges should vanish generally implies a restriction to a
subspace of
the physical Hilbert space. On this subspace the full set of nilpotent
operators,
though non-commuting, could be consistently imposed to vanish.
For $QCD_2$ this means that the vacuum degeneracy
obtained in ref. \cite{8} by solving the cohomology problem in the conformally
invariant sector, described by a G/G topological coset theory
presumes, a priori, that the  ground state of $QCD_2$
lies in the conformally invariant, zero-mass sector of the theory \cite{26}.

We have emphasized the difference in the "currents" involved in the
BRST conditions and the currents generating the symmetries of the original
coupled formulation: With respect to the former, physical states have to be
singlets, whereas these states belong to the irreducible representations
with respect to the latter.

Finally, we have clarified the BRST symmetries underlying the non-abelian
bosonization of free fermions.  The role these symmetries play in assuring
equivalence with the original free fermion dynamics has also been elucidated.

\section{Acknowledgment}

One of the authors (KDR) would like to thank the Physics Department of the
University of Stellenbosch for their kind hospitality.  This work was supported
by a grant from the Foundation of Research development of South Africa.

\end{document}